\documentclass[printer]{aa} % for a paper on 2 columns 
\usepackage{graphicx}
\usepackage{array}
\usepackage{tabularx}
\usepackage{amsmath}
\usepackage{multirow}
\usepackage{chngcntr}
\usepackage{hyperref}
\usepackage{txfonts}
\usepackage{natbib}
\bibpunct{(}{)}{;}{a}{}{,} % to follow the A&A style
\usepackage{units}
\usepackage{comment}
\usepackage{color}
\usepackage[normalem]{ulem}
\begin{document} 
  
  \titlerunning{Rotating bodies in equilibrium with SPH}
  \title{Self-gravitating barotropic equilibrium configurations of rotating bodies with SPH}
   %\title{Recent advances in the smoothed-particle hydrodynamics technique: Building the code SPHYNX}

   \author{D. Garc\'ia-Senz\inst{1,2}, R. M. Cabez\'on\inst{3,4}, J. M. Blanco-Iglesias\inst{1}~and  P. Lor\'en-Aguilar\inst{5}}
\institute{Departament de F\'isica, Universitat Polit\`ecnica de Catalunya, EEBE, Eduard Maristany 16, E-08019 Barcelona, Spain\\
         \email{domingo.garcia@upc.edu}
         \and
        Institut d'Estudis Espacials de Catalunya, Gran Capit\`a 2-4, E-08034 Barcelona, Spain 
        \and        
   Departement Physik, Universit\"at Basel, Klingelbergstrasse 82, 4056 Basel, Switzerland \\
         \email{ruben.cabezon@unibas.ch}
         \and
             Scientific Computing Core, sciCORE, Universit\"at Basel, Klingelbergstrasse, 61, 4056 Basel, Switzerland
    \and
    Department of Physics and Astronomy, University of Exeter, UK
         }

%   \date{Received xxx; accepted xxx}

% \abstract{}{}{}{}{} 
% 5 {} token are mandatory
 
  \abstract
   {Self-gravitational rotating bodies do not have spherically symmetric geometries. The study of physical events appearing in fast-spinning compact stars and accretion disks, as for example those due to localized thermonuclear ignitions in white dwarfs or to the role played by hydrodynamic instabilities in stars and disks, often requires three-dimensional simulations. When the numerical simulations are carried out with the smoothed-particle-hydrodynamics (SPH) technique a critical point arises as to how to build a stable initial model with rotation because there is not a well-established method for that purpose.} %leave it empty if necessary  
   {We want to provide a portable, easy-to-implement methodology for SPH simulations, to procedurally generate physically sound, stable initial conditions for rotating bodies.}
%   % methods heading (mandatory)
   {We explain and validate an easy and versatile novel relaxation method to obtain three-dimensional equilibrium configurations of rotating bodies with SPH. As detailed below, this method is able to relax barotropic, $P(\rho)$, structures in rigid as well as differential rotation. The relaxation procedure strongly relies on the excellent conservation of angular momentum that characterizes the SPH technique.}
   {We have applied our proposal to obtain stable rotating structures of single white dwarfs, compact binaries harboring two white dwarfs, high-density stars approached as a polytropes and accretion disks with rigid as well as differential rotation.} %The SPH results have been validated by comparing the main features (energies, central densities and the polar to equatorial radius ratio) to those obtained with independent, albeit grid-based methods as, for example, the self-consistent field method. Our results show that both methods agree within few percents.}%The relaxed SPH configurations agree with those obtained with the self-consistent field scheme within a few percents.}
   {We present a novel relaxation method to build three-dimensional rotating structures of barotropic bodies using the SPH technique. The method has been successfully applied to a variety of zero-temperature white dwarfs and polytropic self-gravitating structures. Our SPH results have been validated by comparing the main features (energies, central densities and the polar to equatorial radius ratio) to those obtained with independent, albeit grid-based methods, as for example, the self-consistent field method, showing that both methods agree within few percents.}


   \maketitle
%\tableofcontents
\section{Introduction}
\label{sec:introduction}

The smoothed particle hydrodynamics (SPH) technique has been widely used in astrophysics to study highly dynamical, geometrically distorted, and often catastrophic, events such as star formation \citep{spr03}, stellar encounters either from direct collisions \citep{lombardi95} or with finite impact parameter \citep{davies92}. It has been widely used to model the merging of white dwarfs \citep{loren05} and neutron stars \citep{rasio92} leading to gravitational radiation emission. It has also been applied to simulate type Ia \citep{gsenz05,pakmor12} and type II supernova explosions \citep{fryer02,cab18}. It is also an important tool in the modeling of large-scale structure in the Universe \citep{spr05,guedes11}. 

Interestingly, the initial models in several of the scenarios above are simple  barotropic configurations. In the particular case of the merging of two white dwarf (WD) stars, the final outcome could be a type Ia supernova explosion (SNe Ia), being this double-degenerate scenario (DD) one of the favored production channels for these explosions \citep{hil13}. White dwarfs belonging to a compact binary system may have their rotational velocity substantially increased, once the merger sets in, owing to the transfer of angular momentum from the accretion disk. Although it is thought that the accretion of matter from the companion star would lead to differential rotation \citep{yoon05}, the assumption of rigid rotation is easy to implement and very useful to gain insight on many physical problems. Moreover, the rigid rotation hypothesis is adequate in those cases where the transport of angular momentum from the surface to the center of the star is very efficient \citep{piro08} as it could be the case of degenerate objects like WDs. In particular, the fingerprint of the rotation in the thermonuclear explosion of a WD has been studied by \cite{pfan10a, pfan10b} and \cite{gsenz18}. In the latter, the SPH code SPHYNX \citep{cab17} was used to simulate the explosion of a WD with mass $\simeq 1M_\sun$~in fast rigid rotation. In that calculation, the initial model was built using the relaxation scheme proposed in the present work. Another SNe Ia explosion channel involves a single WD accreting mass from a companion, non-degenerate star, through the Roche-Lobe overflow. This second possibility is called the single-degenerate scenario (SD) channel for SNe Ia. Both scenarios involve large amounts of angular momentum, so the question arises on how to adequately model self-gravitating fast-spinning rotators with SPH. 

 Another topic where having good initial rotating models is crucial is the study of the interaction between  the accretion disk around millisecond pulsars with their magnetospheres. Numerical simulations of these scenarios require  the correct modelling of steady accretion disks, characterized by its total mass, angular momentum and particular rotation-law. Although there has been a number of works dealing with the accretion-disk-magnetosphere interaction \citep[and references therein]{parfrey17} none of them were carry out with particle-based codes.  Clearly, it would be very useful to have steady accretion disks with the adequate resolution to perform simulations of these scenarios with SPH.

Unfortunately, there is no general  procedure to build initial conditions for such self-gravitational, rotating equilibrium structures with SPH.  A mapping procedure from an axisymmetric grid of points to a 3D distribution of equal-mass SPH particles was briefly described in  \cite{durisen85} in connection with the study of rapidly rotating n=3/2 polytropes. \cite{smith96} used a similar mapping procedure to simulate the development of a dynamical bar instability  in a spinning polytrope with n=3/2. However, the direct mapping from grid-points to particles usually leads to incomplete equilibrium  because the models presented in those works display excessive numerical noise\footnote{ Nevertheless, the large amount of numerical noise present in the initial models was used by the authors of these works as the seeds of several rotational induced instabilities.}. A numerical scheme to handle disks in presence of pressure gradients was discussed by \cite{ras16} and there are several recipes to approach the initial conditions prior to the dynamic merging of two WDs with SPH codes \citep{dan11}. There is, however, a rich literature concerning grid-based calculations of the equilibrium properties of barotropic, self-gravitating gases in rotation. An iterative algebraic method to obtain axisymmetric equilibrium structures is the self-consistent field (SCF) method developed by \cite{ostr68}. Later, \cite{hach86} successfully applied the SCF method to build zero-temperature spinning white dwarfs. Another approach was proposed by \cite{eri81} in which both, the Poisson equation and the hydrostatic equilibrium equation (which includes the centrifugal force), are used in its integral form and iterated until convergence in a two-dimensional grid of points. Nevertheless, these methods are not directly applicable to build initial models to carry out SPH calculations. This is because the balance between gravity, pressure and centrifugal forces is lost during the mapping procedure from the 2D ordered grid of points to a 3D distribution of particles.    

As far as we know, there is not a public, well-documented, general procedure to obtain self-gravitating  structures in steady rotation with particle-based hydrodynamic codes. In this manuscript, we develop and test an easy, albeit practical,  novel~relaxation scheme to build barotropic, $P(\rho)$, three-dimensional rotating structures in equilibrium under the SPH paradigm.  In our proposal it is not necessary that the initial distribution of particles is matching any prescribed SCF solution prior relaxation. As an example, these relaxed structures can be used as suitable initial conditions to study the explosion of a rotating WD in the SD scenario, as well as the outcome of the WDs merging in the DD scenario. In this work, the relaxation scheme is applied to zero-temperature white dwarfs, to high-density polytropes, and to pseudo-Keplerian disks.  To validate the method we provide quantitative comparisons with reference solutions, like those obtained by \cite{hach86} and \cite{eri85} using iterative time-independent methods such as the  SCF scheme.

In Sec.~\ref{sec:relax} we describe the physical foundations of our proposal. Section \ref{sec:models} presents the application of the method to build three-dimensional, zero-temperature, white dwarfs in rigid rotation. The scheme developed in Sec.~\ref{sec:models} is also applied to the initial setting of two interacting WDs in a compact binary system in Sec.~\ref{sec:DD}. The extension of the scheme to handle differential rotation in white dwarfs and polytropes is explained in Sec.~\ref{sec:differential}. We apply the method to build pseudo-Keplerian accretion disks in equilibrium in Sec.~\ref{sec:disk}. Finally, we present a summary of our findings and the prospects for future work in the conclusions in Sec.~\ref{sec:conclusions}.

\section{Relaxing rotating white dwarfs with SPH}
\label{sec:relax}

 In this section we aim to describe the physical basis of our relaxation scheme to obtain stable rotating structures of barotropic bodies, with particular emphasis in degenerate structures such as cold white dwarfs. Non-rotating equilibrium configurations of WDs can be obtained by relaxing a sample of particles with initial spherical coordinates $(r, \phi, \theta)$. These SPH particles are radially distributed according to the density profile, but randomly in angles $\phi, \theta$. Usually, a damping force is added to the momentum equation so that, after a few sound-crossing times, the sample of particles relaxes to a stable configuration. It is worth noting, however, that as the mass of the WD approaches the Chandrasekhar-mass limit such equilibrium is not perfect and the degenerate star undergoes small radial oscillations. Fortunately, when simulating Type Ia Supernovas, the thermonuclear explosion of a massive WD is so fast that it is enough to keep the equilibrium only during a few sound-crossing times, $t_{sc}$ (typically $t_{sc}\sim 0.46$~s at $\rho\simeq 10^9$~g/cm$^3$).

In the absence of rotation, the structure of the WD after the relaxation process is spherically symmetric and follows the well-known solution of the Lane-Emden (LE) equation.  Thus, for an equation of state (EOS) dominated by a  zero-temperature electron gas  \citep{chandra39},

\begin{equation}
P_e=a~[x(2x^2-3)(x^2+1)^\frac{1}{2}+3 {\mathrm~ sinh}^{-1} x],
\label{e_pressure}
\end{equation}

\begin{equation}
  \rho=b x^3,
\end{equation}

\noindent
where $a=6.00\times 10^{22}$~dynes/cm$^2$, $b=9.82\times 10^5\mu_e$~g/cm$^3$, and $x$~is the Fermi momentum of electrons in relativistic units. The only parameters determining the density and pressure profiles are the mass of the WD and the electron molecular weight $\{M_{WD},~ \mu_e\}$. Unfortunately, there is not a simple description, equivalent to the Lane-Emden equation, but for rotating stars. Assuming that the rotating white dwarf has axisymmetric geometry, its structure is set by the triad $\{M_{WD},~J_{WD},~\mu_e\}$ where $J_{WD}$ is its total angular momentum. For differential rotators, it is necessary to additionally specify the rotational law followed by the angular velocity  $\Omega (s)$, where $s$ is the distance to the rotation axis.

Once these parameters are defined, the maximum density ($\rho_{max}$) and radius ($R_{WD}$) of the resulting configuration will be uniquely determined. It is important to note, that those final values ($\rho_{max}$ and $R_{WD}$) are different to those obtained when rotation is not present, even for the same combination of $\{M_{WD},~ \mu_e\}$. Indeed, once the mass and its composition are fixed, a rotating WD will have a lower $\rho_{max}$ and larger $R_{WD}$ in the rotating plane, than a non-rotating one.

When studying the efficiency of nuclear burning in Type Ia Supernova explosions, the dominant magnitude is the density. Therefore, we fix $\rho_{max}$ between rotating and non-rotating models which, as a consequence, implies a change in the total mass of the WD.

In this work, we have considered the following rotation-law:

 \begin{equation}
     \Omega(s) = \frac{\Omega_c}{\left(1 + \frac{s^2}{R_c^2}\right)^m},
     \label{rotationlaw_1}
 \end{equation}

 \noindent
 where $s$ is the distance to the rotation axis, $R_c$ is a parameter which sets the size of the central core with nearly rigid rotation $\Omega_c$, and $m$ is a parameter linked to the type of rotation (see sections \ref{sec:models} and \ref{sec:differential}). Current choices of $m$ are: $m=0$ (rigid rotation), $m=1/2$, and $m=1$ (shellular). Rigid rotation is also attained for $R_c >> R_{WD}$, independently of $m$.

 Our relaxation method works as follows. Firstly, we choose the values of $\rho_{max}$, $M_{WD}$, and $J_{WD}$ from \cite{hach86} data tables (their Tables 4 and 5), and we take $\mu_e=2$~in the electron zero-temperature equation of state (EOS). We then build an initial model with spherical symmetry (i.e. without rotation) and maximum density, $\rho_{max}$. Such model is obtained after integrating the Lane-Emden equation with inner boundary condition $\rho(r=0)=\rho_{max}$. We note that the total mass of the WD obtained from the LE equation, $M_{WD}^{LE}$, is not the same as $M_{WD}$~of the rotating model in Hachisu's tables for the same central density $\rho_{max}$. The density profile is then mapped to a 3D distribution of equal-mass SPH particles, which ensures that the number density of the particles is reflecting the current density structure at any point. The mass of the particles is then re-scaled by a factor $M_{WD}/M_{WD}^{LE}$, so that the total mass becomes $M_{WD}$. This, of course, changes the central density but only transiently because its value is rapidly restored to $\rho_{max}$~during the relaxation process. Next, a velocity is given to each SPH particle so that the total angular momentum is $J_{WD}$,

 \begin{equation}
     J_{WD}=\sum_b m_b s_b v_b = \left[\sum_b m_b s_b^2 \left(1+\frac{s_b^2}{R_c^2}\right)^{-m}\right]\Omega_c (t),
     \label{momentumcons_1}
 \end{equation}

 \noindent
 from which the instantaneous value of $\Omega_c(t)$ is obtained.

 \begin{equation}
 \Omega_c(t)=\frac{J_{WD}}{\left[\sum_b m_b s_b^2\left(1+\frac{s_b^2}{R_c^2}\right)^{-m}\right]}.
     \label{momentumcons_2}
 \end{equation}

 Note that when $R_c\rightarrow\infty$ the angular velocity becomes $\Omega(t)= J_{WD}/I_{WD}(t)$, where $I_{WD}(t)$ is the time-dependent momentum of inertia of particles around the rotation axis, which is computed at each integration step during the relaxation. For other choices of $R_c$~and $m$~Eq.~(\ref{momentumcons_2}) gives the correct angular velocity as a function of the generalized momentum of inertia of the system, provided $J_{WD}$~is known at t=0. This original and simple recipe is physically sound and robust because it always preserves the total mass and total angular momentum of the system, even for highly deformed axisymmetric structures. 

 Once $\Omega_c(t)$ is known, Eq. (\ref{rotationlaw_1}) gives $\Omega(s,t)$, so that the centripetal acceleration of each particle $\bf{\Omega}\times(\bf{\Omega}\times \bf {r})$ is obtained.
 The particle distribution is henceforth relaxed in a co-moving reference frame.  We then let the system freely evolve with the SPH code until equilibrium.  
 
 To relax the system and dissipate the spurious numerical noise stored in the velocity field  the velocities are periodically set to zero (in the co-moving frame of reference). For example, all models in Table \ref{table_1} with N$_p=5\times 10^5$~particles were obtained by setting $v=0$~every $\Delta t \simeq t_{sc}/3$, being $t_{sc}$~the sound-crossing time ($t_{sc}\simeq 0.46$~s in models with central density $\rho_c=10^9$~g~cm$^{-3}$) until time $t\simeq 5t_{sc}/3$. Afterwards the velocities were set to zero every $\Delta t\simeq 0.8 t_{sc}$~s. Such simple recipe works well and can be adapted to handle different particle numbers and densities. After several sound crossing times, the final equilibrium configuration is attained and only a residual numerical noise remains.  To decide when the particle distribution has converged to a stable, time-independent configuration, each calculated model has to fulfill two criteria: 1) that its central density and equatorial radius become constant (except small oscillations with amplitude 1-2\% around their average value), 2) that the central density, the polar and equatorial radius, and the total kinetic, internal and gravitational energies do remain below the percent above when the system is letting free to evolve in the current inertial frame (i.e. without re-setting the velocities to zero).

%\section{Calculated models}
\section{Isolated WDs with rigid rotation}
\label{sec:models}

 As a first application we focus on the evolution towards equilibrium of models calculated assuming rigid rotation for both, single WDs and double degenerate stars in compact binaries. Some insights about how to handle differential rotation, obeying the  rotation law given by Eq.~(\ref{rotationlaw_1}) with $m\ne 0$ are provided in Sec.~\ref{sec:differential}. The simulations presented here, as well as in the remaining sections, were carried out with the hydrodynamic code SPHYNX, using the electron zero-temperature EOS given by Eq.~(\ref{e_pressure}) with $\mu_e=2$. The calculations were performed with the default values of several parameters, as for example: $\alpha=4/3$~in the artificial viscosity (AV), with the  Balsara limiters \citep{balsara95} to the AV turned on. The gravitational force is approached by a multipole expansion up to quadrupole terms, with the tolerance parameter $\theta$~\citep{hernquist87}, set to $\theta =0.6$. An exponent  parameter, $p$ ($0\le p \le 1$), allows to choose among different volume elements (VE) \citep{cab17}. The present simulations were obtained with $p=0$, which is equivalent to the common choice  $VE=m/\rho$.

A cautionary remark concerning rigid fast rotation close to  Keplerian values, is necessary here. For these models, we observed that the combination of solid rigid rotation and a high rotational speed led to a harmful feedback between the centrifugal, gravitational, and pressure forces at the equator surface: let's assume that there is a particle with a slight excess of angular velocity which displaces it farther in radius, thus increasing the centrifugal force (but weakening gravity). Such particle will move farther out, which again increases its the centrifugal force and so on. Even though such feedback was only affecting a handful of surface particles located on the equator it was sufficient to spoil the convergence. If we want to make quantitative comparisons between our results and those by the SCF method for perfect rigid rotators, a solution is to (artificially) reduce a bit the gradient of pressure at the surface, so that the feedback above is broken, even in the presence of low amounts of numerical noise. A simple way to do that and keep the relaxation stable is to reduce the pressure in the low-density regions of the WDs by multiplying the electron pressure, Eq.~(\ref{e_pressure}), by a density cut-off,

\begin{equation}
P =
\begin{cases}
P_e & \rho > \rho_{crit}\\
P_e\times\frac{\rho}{\rho_{crit}} & \rho \le \rho_{crit}
\end{cases}
\label{pcutoff}
\end{equation}

\noindent
where $P$ is the pressure used in the calculations and $\rho_{crit}=\alpha~\rho_{max}$, being $\rho_{max}$ the maximum density, which in the models shown in Table \ref{table_1}~is attained at the center of the configuration.  We empirically found that  $\alpha=5\times 10^{-4}$ works well in almost all the rigidly rotating models shown in Table~\ref{table_1}. Slightly lower values, $\alpha=2.5\times 10^{-4}$~and $\alpha =2\times 10^{-4}$, were used to relax the more massive models $A_{21}$ and $A_{22}$, to reduce the error in the equatorial radius. Other magnitudes, such as the central density and total kinetic, internal and gravitational energies are rather insensitive to the precise value of $\alpha$. We stress again that such density cut-off is merely a way to facilitate the convergence of the relaxation. It affects a negligible amount of mass, and it makes possible a direct comparison with the results obtained with the SCF method for rigid rotators. In real calculations, however, it is preferable not to consider absolute rigid rotation and allow for some amount of differential rotation at the external layers. In that case, introducing the above density cut-off becomes unnecessary.

% PROVISIONAL. FIGURE 1.  EVOLUTION OF CENTRAL DENSITY and OMEGA. COLORMAP.  
\begin{figure*}
\centerline{\includegraphics[width=\textwidth]{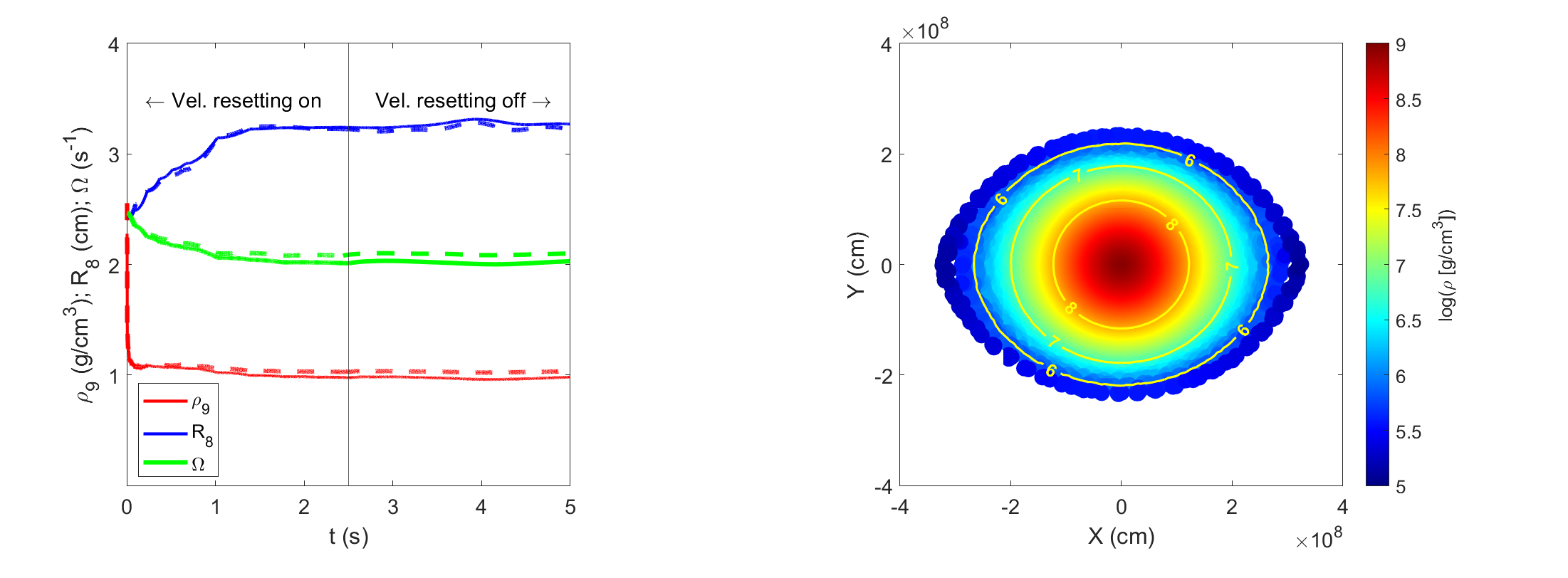}
% where an .eps filename suffix will be assumed under latex,
% and a .pdf suffix will be assumed for pdflatex
}
\caption{Simulation results of models $A_{21}$ and $A_{22}$ of Table~\ref{table_1}. Left: evolution of central density (in $10^9$~g/cm$^3$), radius (in $10^8$~cm), and angular velocity $\Omega$.  Solid lines correspond to model $A_{21}$ ($5\times 10^5$ particles) and dashed lines to model $A_{22}$ ($2\times 10^6$ particles). Once the periodic resetting to zero of the velocities is removed at $t=2.5$~s, the central density remains stable during several sound-crossing times ($t_{sc}\simeq 0.4$~s). Right: density colormap and isodensity contours of a 2D meridional slice of the rotating WD at $t=2.5$~s. The slice has a thickness of four times the local smoothing length (4h), which roughly represents $10$\% of the total number of particles. Making cuts with the local value of $h$~ensures a similar amount of particles at  any region of the color-map.}
\label{fig:rigid}
\end{figure*}

\begin{table*}
%%% increase table row spacing, adjust to taste
\renewcommand{\arraystretch}{1.4}
\caption{Models with rigid ($m=0$ in Eq.~\ref{rotationlaw_1}) and differential ($m=1/2$) rotation of zero-temperature white dwarfs. Models A are the SPH calculations performed with SPHYNX. Models H are the SCF calculations by \cite{hach86}. The columns are: rotation-law (m), mass of the star (M$_{WD}$), total angular momentum (J$_{WD}$), number of particles (N$_p$), total kinetic, internal, and gravitational energies (E$_{k}$, E$_I$, E$_{G}$), maximum density ($\rho_{max}$), minimum smoothing-length (h$_{min}$), equatorial radius (R$_{eq}$), and polar to equator radius ratio (F). Energies and densities are in c.g.s. units.}
\label{table_1}
\begin{center}
%%% Some packages, such as MDW tools, offer better commands for making tables
%%% than the plain LaTeX2e tabular which is used here.
\begin{tabular}{|c|c|c|c|c|c|c|c|c|c|c|c|}
\hline
%\bf{Model} & \multicolumn{3}{|c|}{\bf{Table Column Head}}\\
\bf{}
%\cline{2-4}
&\multirow{2}{*}{m} &{M$_{wd}$} & J$_{wd}$&{N$_p$}&{E$_{k}$} & {E$_{I}$}& {-E$_{G}$} &$\rho_{max}$&h$_{min}$&$R_{eq}$&\multirow{2}{*}{F} \\
 & & M$_\odot$&$\times10^{50}$&$\times10^{6}$& $\times10^{50}$& $\times 10^6$&$\times10^{50}$ & $\times10^6$& km& km& \\
\hline
\hline
A$_1$& \multirow{2}{*}{0}& \multirow{2}{*}{1.35} &\multirow{2}{*}{0}  &0.5&0 &18.1 &22.6 &1004&26&2360 &1  \\
\cline{1-1}\cline{5-12}
H$_1$ & &  &  & -&0& 18.1& 22.6& 1000&-&2460 &1 \\
\hline
\hline
A$_{21}$ &\multirow{3}{*}{0}&\multirow{3}{*}{1.44} &\multirow{3}{*}{0.522} &0.5&0.525 &18.7 &24.4 &977 &25& 3240&0.716\\
\cline{1-1}\cline{5-12}
A$_{22}$ & & & &2.0&0.543 &19.1 &24.9 &1028 &16& 3230&0.710\\
\cline{1-1}\cline{5-12}
H$_2$ & & & &-& 0.537&18.9 &24.6 &1000&- &3500&0.667\\
\hline
\hline
A$_3$ &\multirow{2}{*}{0}&\multirow{2}{*}{1.28} &\multirow{2}{*}{0.745}&0.5&0.313 & 6.32& 9.78& 102&54&5820 &0.687\\
\cline{1-1}\cline{5-12}
H$_3$ & & & &- &0.313 &6.25&9.69 & 100&-& 6040&0.667\\
\hline
\hline
A$_4$ &\multirow{2}{*}{0}&\multirow{2}{*}{0.908} &\multirow{2}{*}{0.707}&0.5 & 0.118&1.43 &2.67 &9.91&106 &9470 &0.678\\
\cline{1-1}\cline{5-12}
H$_4$& & & &- &0.119 &1.43 &2.69&10&- &9720 &0.667\\
\hline
\hline
A$_5$ &\multirow{2}{*}{0}&\multirow{2}{*}{0.674} &\multirow{2}{*}{0.543}&0.5 &0.056 & 0.564& 1.134&3.13&138 &12000 &0.663 \\
\cline{1-1}\cline{5-12}
H$_5$ & & & &- & 0.056& 0.553&1.14 &3.16&- & 12100&0.667\\
\hline
\hline
A$_6$ &\multirow{2}{*}{1/2}& \multirow{2}{*}{1.65} &\multirow{2}{*}{0.993}&0.5 & 1.92 &22.9 &32.3 &970&27&2680 &0.698  \\
\cline{1-1}\cline{5-12}
H$_6$& & & &- & 1.96& 23.3& 32.8&1000&-&2720 & 0.667\\
\hline
\hline
A$_7$ &\multirow{2}{*}{1/2}&\multirow{2}{*}{1.99} &\multirow{2}{*}{1.86}&0.5 &4.78 &29.1 &45.5 &980 &28& 2900&0.510\\
\cline{1-1}\cline{5-12}
H$_7$& & & &-& 4.87&29.6 &46.1 &1000 &-&2940&0.500\\
\hline
%\hline
%\hline
%$_1$& 0& 1.35 &0  &0.5&0 &18.1 &22.6 &1004&26&2360 &1  \\
%\hline
%A$_{21}$ &0&1.44 &0.522 &0.5&0.525 &18.7 &24.4 &977 &25& 3240&0.716\\
%\hline
%A$_{22}$ &0&1.44 &0.522 &2.0&0.543 &19.1 &24.9 &1028 &16& 3230&0.710\\
%\hline
%A$_3$ &0&1.28 &0.745&0.5&0.313 & 6.32& 9.78& 102&54&5820 &0.687\\
%\hline
%A$_4$ &0&0.908 &0.707&0.5 & 0.118&1.43 &2.67 &9.91&106 &9470 &0.678\\
%\hline
%A$_5$ &0&0.674 &0.543&0.5 &0.056 & 0.564& 1.134&3.13&138 &12000 &0.663 \\
%\hline
%A$_6$ &1/2& 1.65 &0.993&0.5 & 1.92 &22.9 &32.3 &970&27&2680 &0.698  \\
%\hline
%A$_7$ &1/2&1.99 &1.86&0.5 &4.78 &29.1 &45.5 &980 &28& 2900&0.510\\
%\hline
%\hline
%H$_1$ &0& 1.35 &0 & -&0& 18.1& 22.6& 1000&-&2460 &1 \\
%\hline
%H$_2$ &0&1.44 & 0.522&-& 0.537&18.9 &24.6 &1000&- &3500&0.667\\
%\hline
%H$_3$ &0&1.28 &0.745&- &0.313 &6.25&9.69 & 100&-& 6040&0.667\\
%\hline
%H$_4$&0&0.908 &0.707&- &0.119 &1.43 &2.69&10&- &9720 &0.667\\
%\hline
%H$_5$ &0&0.674 &0.543&- & 0.056& 0.553&1.14 &3.16&- & 12100&0.667\\
%\hline
%H$_6$&1/2 & 1.65 &0.993&- & 1.96& 23.3& 32.8&1000&-&2720 & 0.667\\
%\hline
%H$_7$&1/2 &1.99 & 1.86&-& 4.87&29.6 &46.1 &1000 &-&2940&0.500\\
%\hline

\end{tabular}
\end{center}
\end{table*}

%White dwarfs belonging to a compact binary system may have its rotational velocity substantially increased\rc{, once the merger sets in,} owing to the transference of angular momentum from the accretion disk. Although it is thought that the accretion of matter from the companion star would lead to differential rotation \citep{yoon05}, the assumption of rigid rotation is easy to implement and very useful to gain insight on many physical problems. Moreover, the rigid rotation hypothesis is adequate in those cases where the transport of angular momentum from the surface to the center of the star is very efficient \citep{piro08} as it could be the case of degenerate objects like WDs.
%In particular,  the fingerprint of the rotation in the thermonuclear explosion of a WD has been studied by \citep{pfan10a, pfan10b} and \citep{gsenz18}. In the latter, the code SPHYNX was used to simulate the explosion of a WD with mass $\simeq 1M_\sun$~in fast rigid rotation. In that calculation, the initial model was built using the relaxation scheme proposed in the present work.  

%  FIGURE 2. SNAPSHOTS OF RELAXATION DD. 
\begin{figure*}
\centerline{\includegraphics[width=\textwidth]{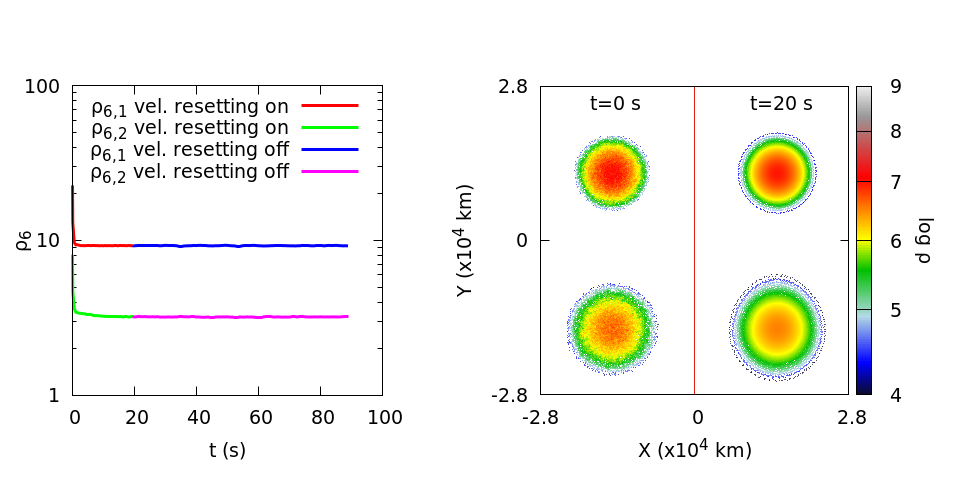}}
% where an .eps filename suffix will be assumed under latex,
% and a .pdf suffix will be assumed for pdflatex
\label{fig:DD_case}
\hfil
\caption{Initial setting of the DD scenario. Left: Evolution of the central density of both WDs during the relaxation ($t\le 20$~s) and the free evolving ($t\ge 20$~s) stages. The orbital period is $P\simeq 70$~s. Right: Slice in the orbital plane depicting the density colormap of each white dwarf at $t=0$~s (initial spherically symmetric configuration) and $t=20$~s (final relaxed model).  The slice has a thickness of four times the local smoothing-length value (4h). The center of mass of both configurations is located at (0,0) km, but both snapshots have been shifted 14,000~km to the left and to the right to avoid the superposition of the images.}
\label{fig:DD_case}
\end{figure*}

 Here we carry out the relaxation of several WDs with different \{M$_{WD}, J_{WD}\}$. An estimation of the accuracy of the resulting equilibrium configurations is done by comparing our results to those by \cite{hach86}, being the latter obtained using the SCF method.

 We provide a representative example of the evolution towards equilibrium of one of our rotating models in the left panel of Fig.~\ref{fig:rigid}, which depicts the evolution of the central density $\rho_{c}$~($\rho_c=\rho_{max}$~in these models), equatorial radius, $R_{eq}$~and angular velocity $\Omega$. As we can see, the central density and angular velocity  evolve in a similar manner. They start from relatively high values, decrease fast during a couple of tenths of a second, and stabilize at $t\simeq 1$~s. The equatorial radius follows the opposite trend, it is low at the beginning, then it rises fast to asymptotically stabilize at $t\simeq 2.0$~s. From Fig.~\ref{fig:rigid} it is obvious that the characteristic relaxation times of $\rho_c$~and $R_{eq}$~are quite different. To check that the star was stable we  stop resetting the velocities to zero at $t=2.5$~s, so that the system evolves freely in the current inertial frame of reference.  The only source of dissipation being the artificial viscosity (AV) term, as given by \cite{mon97}, including the Balsara limiters. As it can be seen, the central density remained stable during at least $\Delta t\simeq 3$~s ($\simeq$~7 times the sound-crossing time).  We show a color-map slice of the density of model $A_{21}$~at $t=2.5$~s in Fig.~\ref{fig:rigid} (right). The white dwarf is neatly oblated with a polar to equator radius $F=\frac{R_p}{R_{eq}}= 0.716$~which is $\simeq 7.3\%$~larger than that given by the SCF method. This is, however, the largest discrepancy found across all the calculated models in Table~\ref{table_1}. The differences with respect the results by Hachisu remain, for the most part, below  $4\%$. 

Table~\ref{table_1} summarizes the relevant information of the calculated models. Models A refer to the SPHYNX calculations and H refer to the SCF models by Hachisu. Models A$_1$ and H$_1$ are non-rotating, with a mass close to the Chandrasekhar-mass limit. As we can see, the fit is excellent. The larger discrepancy, $\simeq 4\%$, is in the radius of the configuration. Actually, this is something expected because a white dwarf nearing the Chandrasekhar-mass limit has a not well-defined scale-length. We thus expect that the larger differences with respect the SCF method affect the equatorial radius and the oblateness of the WDs at central densities $\rho_{9}\ge 1$~.  Model $A_{22}$~in Table \ref{table_1} is the same as $A_{21}$~ but calculated with four times more particles. It leads to a stable model with similar relative errors in the central density, equatorial to polar radius and total energies, with respect to those of the reference model $H_2$. Therefore, our relaxation method is able to match the results by Hachisu in a wide range of stellar masses,  $0.67 M_{\odot} \le M_{WD}\le 1.44 M_{\odot}$. The lower mass is close to that of a standard WD and the higher mass is actually at the Chandrasekhar-mass limit of a non-rotating white dwarf. The case of a super-Chandrasekhar mass white dwarf stabilized by rotation is discussed in Sec. \ref{subsec:differentialWD}.

%\section {Initial models of compact double degenerate binaries.}
\section {WDs in a double degenerate binary system}
\label{sec:DD}

A straightforward and timely extension of the relaxation procedure described above can be used to generate suitable initial models to study the DD production channel for SNe Ia. In the DD channel two WDs, settled in a compact orbit, get closer because of gravitational-wave radiative losses. At some point the gravitational pulling from the more massive WD breaks the lighter compact star and an accretion disk around the surviving WD forms. The further accretion of the debris would eventually provoke the explosion of the initially more massive white dwarf \citep{hil13}.

A key technical point of the simulations of the DD scenario is how to set the initial conditions immediately prior the merging. The nowadays accepted standard procedure involves a two step process \citep{ross04, dan11}. In the first place, both stars are relaxed in isolation. Then, both stars are placed in a wide enough binary orbit in order to prevent any immediate mass transfer episodes. Subsequently, the binary system is evolved in the co-rotating frame where an artificial acceleration term is introduced in order to continuously shrink the orbital distance between both white dwarfs. Orbital distance will be decreased at a sufficiently slow rate so the stars remain relaxed at all times, and slowly and continuously deform without introducing any spurious oscillations. The relaxation process will be finished when the secondary star starts to overflow its Roche Lobe. This method is, however, computationally expensive and somehow artificial because both WDs need to be relaxed all the time along the path from the initially detached position until they reach the onset of the merger (usually achieved through the introduction of additional artificial dissipative terms).

A somehow more elegant and fast procedure is to relax both stars taking into account their angular momentum once they are already settled in a close orbit just prior the merging, thus avoiding the intermediate relaxation stages. To do that we first set the orbit parameters so that the gravitational pulling onto the surface of the less massive WD becomes a sizable fraction ($\beta$) of its own gravity. That constraint leads to the following expression for the distance,

\begin{equation}
      D_{1,2}=R_2\times\left(1+\sqrt{\frac{M_{WD1}}{\beta~M_{WD2}}}~\right).
      \label{distbinary}
\end{equation}

\noindent
where $M_{WD1}$ and $M_{WD2}$ are the masses of the more massive and lighter components, respectively. Once $D_{1,2}$ is known, and assuming a circular orbit,  we calculate the velocity of the center of mass of each star with respect an inertial reference frame located at rest at the center of mass of the binary system. The total angular momentum of the binary system $J_{sys}$ is afterwards calculated, so that we can benefit from the scheme developed above, in Sections \ref{sec:relax} and \ref{sec:models}.

\begin{table*}
%%% increase table row spacing, adjust to taste
\renewcommand{\arraystretch}{1.3}
\caption{Main features of the Double Degenerate models. The central densities $\rho_{max1}$~and $\rho_{max2}$ are the values at the end of the relaxation period. The number of particles allocated in each star (N$_{p1}$, N$_{p2}$) is also  provided, as well as the minimum value of the smoothing lengths (h$_{min1}$, h$_{min2}$).}
\label{table_2}
\begin{center}
%%% Some packages, such as MDW tools, offer better commands for making tables
%%% than the plain LaTeX2e tabular which is used here.
\begin{tabular}{|c|c|c|c|c|c|c|c|c|c|c|c|c|}
\hline
%\bf{Model} & \multicolumn{3}{|c|}{\bf{Table Column Head}}\\
%\cline{2-4}
 & {M$_{wd1}$} &{M$_{wd2}$}& D$_{1,2}$ & J$_{sys}$ & \multirow{2}{*}{$\frac{1}{\beta}$} &N$_{p1}$&N$_{p2}$& $\rho_{max1}$ & $\rho_{max2}$&h$_{min1}$&h$_{min2}$ & P \\
 & M$_\odot$ & M$_\odot$&$10^9$~cm & $\times10^{50}$~erg.s & &$\times 10^6$&$\times 10^6$&$\times10^{6}$~g.cm$^{-3}$ &$\times10^6$~g/cm$^3$&km&km&s \\
\hline
\hline
DD$_1$ & 0.606 & 0.606 & 2.589 & 4.21448 &4.0&0.25&0.25 &3.17 & 3.17&150&150& 65 \\
\hline
DD$_2$ &0.796 &0.606 & 2.842 &5.25972 &4.0&0.328&0.25 &9.22 & 3.21&100&151& 70 \\
\hline
\end{tabular}
\end{center}
\end{table*}

As an example, we have calculated the stable initial configurations in two cases. We have first considered a system of twin white dwarfs with canonical masses $M_{WD1}=M_{WD2}=0.606$~M$_{\odot}$ and parameter $\beta= 1/4$. The second case is for $M_{WD1}=0.796$~M$_\odot$, $M_{WD2}=0.606$~M$_\odot$, and the same value of $\beta$. The center of the compact stars is supposed to move in circles around the center of mass of the binary system, with  angular velocity $\Omega$. The total angular momentum of the system is $J_{sys}= J_{orb} + J_{spin}$. In this work we focus on pairs of white dwarfs which are tidally locked so that $\Omega_{spin}=\Omega_{orb}=\Omega$.
Therefore, once $\Omega$~is deduced from 

\begin{equation}
    \Omega= \sqrt{G~\frac{(M_{WD1}+M_{WD2})}{D_{1,2}^3}}\,
    \label{Omega}
\end{equation}

\noindent
 the total angular momentum $J_{sys}=J_{orb} + J_{spin}$~is easily obtained.\footnote{ This is not the unique way to make a reasonable guess of $J_{sys}$ before relaxation. Another possibility to estimate $J_{sys}$ is to use the modified Kepler's law by \cite{lai1994} which results from considering deformed ellipsoids. In any case, what really matters is that, once it has converged, the DD system has a structure compatible with the initial choice of J$_{sys}$.} The post-relaxed binary configuration in rigid rotation will be determined by  \{$M_{WD1}, M_{WD2}, J_{sys}$\}. The EOS was that of a zero-temperature electron gas given by Eq. (\ref{e_pressure}). A summary of the parameters used in these simulations is given in Table \ref{table_2}.

The evolution of the central densities of both WDs in model DD$_2$~is shown in Fig.~\ref{fig:DD_case} (left). The initial values of central density for both stars are considerably higher at the beginning of the relaxation process, as expected from spherically symmetric initial models. As the simulation proceeds, the system rapidly achieves a stable configuration with lower densities due to the influence of both rotational effects and tidal forces as well as to the re-ordering of the particles. This stable configuration is reached very fast ($\sim 1$~s for the most massive star and $\sim 5$~s for the other). At $t=20$~the periodic reset to zero of the velocities was turned-off,  without any appreciable effect in the evolution of both stars. We also provide the equilibrium configuration of the WDs in the same figure (right panel), obtained in the co-moving non-inertial frame of reference located at the center of mass. Each dot represents an SPH particle within a thin cut along the plane $Z=0$. Color shows the logarithm of density. Comparing both snapshots, it is clear that the initially spherical random particle distribution, reaches a more ordered distribution, where the least massive star is asymmetrically elongated in the direction of the more massive, as expected. After the relaxation, we checked that the final structure remains stable during at least one complete orbit in the inertial frame. Although the final binary configuration is stable, it is at the verge of the Roche-Lobe overflow. To trigger the catastrophic merging of the WDs it is enough to slightly shrink the distance $D_{1,2}$.

\section{Differential rotation}
\label{sec:differential}
It is feasible to apply the proposed relaxation method to handle differential rotation. We analyze here the impact of taking $m\ne 0$~in Eq. \ref{rotationlaw_1} with different values of the parameter $R_c$. We first discuss the case of  rotating polytropes of index $n=3/2$~with central densities and sizes close to those characterizing neutron stars (NS). The case of a zero-temperature rotating white dwarf with differential rotation is discussed later. In all cases, the results were checked with well-known existing solutions obtained with other independent methods. 

\subsection{Case 1: differentially rotating high density polytropes}

\label{subsec:differentialNS}

% FIGURE 3.  
\begin{figure*}
\centerline{\includegraphics[width=0.85\textwidth]{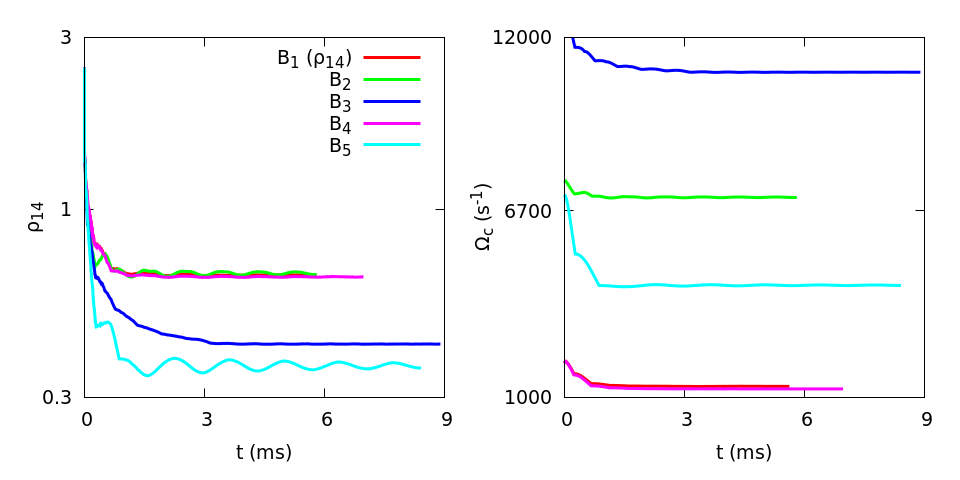}}
% where an .eps filename suffix will be assumed under latex,
% and a .pdf suffix will be assumed for pdflatex
\hfil
\caption{Approach to equilibrium of models described in Table \ref{table_3} in function of time. Density (left panel) and $\Omega_c$ (right panel).}
\label{fig:NSrelax}
\end{figure*}

\begin{table*}
\renewcommand{\arraystretch}{1.5}

\caption{Main features of the relaxed high-density rotating polytropes ($\rho_{max}= 10^{14}$~g/cm$^3$~prior relaxation) calculated with $\gamma=5/3$~and mass $M=2M_\sun$. Models B$_n$~refer to the 3D-SPH calculations and EM$_n$~are the reference results by \cite{eri85}. Columns are: model name, exponent in the rotation law of Eq.~\ref{rotationlaw_1} ($m$) , $A=R_c/R_{NS}$, total angular momentum (J), maximum density ($\rho_{max}$), dimensionless angular momentum (j),  number of SPH particles (N$_p$), minimum smoothing length (h$_{min}$), internal to gravitational energy ratio, kinetic to gravitational energy ratio, total energy conservation, Virial Theorem, central angular velocity, and polar to equatorial radius ratio. The symbol $-$ indicates that the corresponding magnitude was not available in the paper by EM.}
\label{table_3}
\begin{center}
%%% Some packages, such as MDW tools, offer better commands for making tables
%%% than the plain LaTeX2e tabular which is used here.
\begin{tabular}{|c|c|c|c|c|c|c|c|c|c|c|c|c|c|}
\hline
%\bf{Model} & \multicolumn{3}{|c|}{\bf{Table Column Head}}\\
%\cline{2-4}
  &\multirow{2}{*}{m} & \multirow{2}{*}{A} & J & $\rho_{max}$& \multirow{2}{*}{j}&N$_p$ &h$_{min}$& \multirow{2}{*}{$E_I/E_G$}& \multirow{2}{*}{$E_k/E_G$}& $E_T/E_0$ & V.T & $\Omega_c$&\multirow{2}{*}{F} \\
  & & &$\times 10^{49}$~erg.s&$\times 10^{14}$~g/cm$^3$& & $\times 10^6$ &km & & &$\times 10^{-5}$& $\times 10^{-3}$ &s$^{-1}$& \\
\hline
\hline
B$_1$ & \multirow{2}{*}{1} & \multirow{2}{*}{2} & 2.3354 & 0.656 & 0.05129&1.0&0.55&0.429& 0.0685& -7.822 & 4.4& 1336& 0.68 \\
\cline{1-1}\cline{4-13}
EM$_1$ &  &  & - & - & 0.05129&-&- &0.432& 0.0679& -7.759&0.43 &- & 0.68  \\
\hline
\hline
B$_2$ &\multirow{2}{*}{1} & \multirow{2}{*}{0.2} & 1.2946 &0.664 &0.02849&1.0&0.55&0.469 & 0.0385& -2.583 & 3.0& 7116&0.76 \\
\cline{1-1}\cline{4-13}
EM$_2$ & & & - & - &  0.02849&-&- &0.463 & 0.0377& -2.633&0.54 &- & 0.77 \\
\hline
\hline
B$_3$ & \multirow{2}{*}{1} & \multirow{2}{*}{0.2} & 2.1707 & 0.422 &0.04429&1.0&0.63 &0.413& 0.0852& -6.275& 3.2& 10937& 0.54 \\
\cline{1-1}\cline{4-13}
EM$_3$ & &  & - & - & 0.04471&-&- &0.419 & 0.0808& -6.280& 0.49& -& 0.55 \\
\hline
\hline
B$_4$ & \multirow{2}{*}{$1/2$} &\multirow{2}{*}{2} & 2.4306 &0.647 &0.05326&1.0&0.55  & 0.427& 0.0747&-8.320 &2.7 &1260&0.60   \\
\cline{1-1}\cline{4-13}
EM$_4$ & & & - & - & 0.05326&-&-  &0.427 & 0.0731& -8.386&0.40&- & 0.65  \\
\hline
\hline
B$_5$ & \multirow{2}{*}{$1/2$} &\multirow{2}{*}{0.2} & 3.0946 &0.368 &0.06173&1.0&0.67 &0.388 & 0.1110&-12.0 &2.2 &4429 &0.54\\
\cline{1-1}\cline{4-13}
EM$_5$ & & & - & - & 0.06173&-&- &0.396 & 0.1037& -11.8 &0.45 & - & 0.57 \\
\hline
%\hline
%\hline
%B$_1$ & 1 & 2 & 2.3354 & 0.656 & 0.05129&1.0&0.55&0.429& 0.0685& -7.822 & 4.4& 1336& 0.68 \\
%\hline
%B$_2$ &1 & 0.2 & 1.2946 &0.664 &0.02849&1.0&0.55&0.469 & 0.0385& -2.583 & 3.0& 7116&0.76 \\
%\hline
%B$_3$ & 1 & 0.2 & 2.1707 & 0.422 &0.04429&1.0&0.63 &0.413& 0.0852& -6.275& 3.2& 10937& 0.54 \\
%\hline
%B$_4$ & $1/2$ &2 & 2.4306 &0.647 &0.05326&1.0&0.55  & 0.427& 0.0747&-8.320 &2.7 &1260&0.60   \\
%\hline
%B$_5$ & $1/2$ &0.2 & 3.0946 &0.368 &0.06173&1.0&0.67 &0.388 & 0.1110&-12.0 &2.2 &4429 &0.54\\
%\hline
%\hline
%EM$_1$ & 1 & 2 & - & - & 0.05129&-&- &0.432& 0.0679& -7.759&0.43 &- & 0.68  \\
%\hline
%EM$_2$ & 1 &0.2 & - & - &  0.02849&-&- &0.463 & 0.0377& -2.633&0.54 &- & 0.77 \\
%\hline
%EM$_3$ & 1 & 0.2 & - & - & 0.04471&-&- &0.419 & 0.0808& -6.280& 0.49& -& 0.55 \\
%\hline
%EM$_4$ &1/2 &2 & - & - & 0.05326&-&-  &0.427 & 0.0731& -8.386&0.40&- & 0.65  \\
%\hline
%EM$_5$ &1/2 &0.2 & - & - & 0.06173&-&- &0.396 & 0.1037& -11.8 &0.45 & - & 0.57 \\
%\hline

\end{tabular}
\end{center}
\end{table*}

% FIGURE 4.  
\begin{figure*}
\centerline{\includegraphics[width=\textwidth]{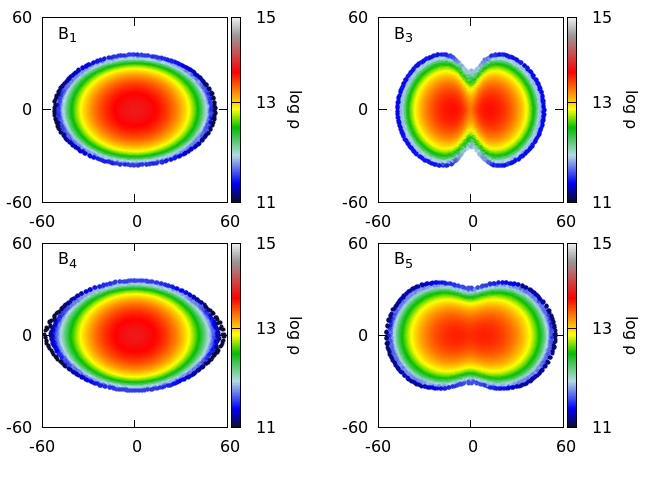}}
% where an .eps filename suffix will be assumed under latex,
% and a .pdf suffix will be assumed for pdflatex
\hfil
\caption{Colormap of density in a meridional cut (with thickness 4h) of several models as described in Table \ref{table_3}. }
\label{fig:NScolormap}
\end{figure*}

To check the ability, as well as the limits, of our relaxation scheme to handle non-rigid rotation we chose a polytropic relation

\begin{equation}
P=K\rho^\gamma
\label{polytropicEOS}
\end{equation}

\noindent with $\gamma=5/3$. In non-rotating models such value of $\gamma=5/3$ produces very stable spherically symmetric objects, whose structure is obtained after solving the Lane-Endem equation with $n=1/(\gamma-1)=3/2$. 

Our initial, pre-relaxed, model is a spherically symmetric object with mass $M=2M_\sun$~and central density $\rho_c=10^{14}$~g/cm$^3$. Such combination sets the polytropic constant in Eq.~(\ref{polytropicEOS}) to $K=1.72\times 10^{10}$. The integration of the LE equation leads to an object of radius $R_{NS}\simeq 38$~km, roughly compatible with a NS size.  Some amount of total angular momentum $J$~was then added to the polytrope and the structure relaxed with the method explained in Section \ref{sec:relax}. The angular momentum  $J$~was chosen so that the square of the dimensionless angular momentum $j$,

\begin{equation}
    j^2 = \frac{J^2}{4 \pi G M^{\frac{10}{3}}\rho_{max}^{-\frac{1}{3}}}
    \label{j_angmom}
\end{equation}

 \noindent matches the values~in Tables 1 and 2 by Eriguchi \& Mueller (EM) \citep{eri85} so that meaningful comparisons with their results can be done. Note, however,  that the value of $\rho_{max}$ in Eq.~(\ref{j_angmom}) initially differs from that in EM because we take $\rho_{max}=\rho_c$ of the spherically symmetric polytrope, instead of the maximum density of the rotating structure used by EM. Such initial choice of $\rho_{max}$ is motivated by the fact that we do not know the true value of the density of the rotating structure prior relaxation. Nevertheless, this does not pose a problem because the value of $j$ can be recalculated with the value of $\rho_{max}$ obtained once the relaxation has ended. Such new value of $j$ (6th column in Table~\ref{table_3}) was the one used to check our results with those brought in the EM tables.   
 
 We calculated three models using $j$-constant ($m=1$) and two more with $v$-constant ($m=1/2$) rotational laws. In each case we consider $R_c= AR_{NS}$ with $A=2$ and $A=0.2$ in Eq.~(\ref{rotationlaw_1}), respectively. The case $A=2$ is actually close to rigid rotation but the case $A=0.2$~is representative of models with large differential rotation. Table~\ref{table_3} shows a summary of the results, where the meaning of the columns is as follows: third column (J) is the total angular momentum, $\rho_{max}$ is the maximum density of the rotating body (not necessarily achieved at the center), $E_k$, $E_I$, and $E_G$ are the total kinetic, internal and gravitational energies, respectively, and $E_0$ is a normalization energy defined by 
 
 \begin{equation}
 E_0=\frac{(4\pi G)^2 M^5}{J^2}.
 \label{energy_0}
 \end{equation}
 
 The symbol V.T refers to the virial theorem defined as V.T~$= \vert(2E_k+E_G+3(\gamma-1)E_I)/E_G\vert$, $\Omega_c$ is the angular velocity at the center of the polytrope and, finally, $F=R_p/R_{eq}$ is the polar to equatorial radius. 
 
 As shown in Table~\ref{table_3}, the global energies and the polar to equator ratio agree to the results by EM within a few percents. Nevertheless, models B$_n$~do not fulfill the virial theorem so well as in the EM$_n$~calculations. There is almost a factor ten difference between both estimations. Such discrepancy arises from the very different approach to the structure equilibrium of the rotating polytropes. In the case of the SPH calculation the equilibrium is approached dynamically in full three dimensions while in the EM calculation the equilibrium equations are solved by means of a time-independent iterative procedure in a fixed two-dimensional grid of points.
 
 Figure~\ref{fig:NSrelax} depicts the evolution of $\rho_{max}$~(left panel) and $\Omega_c$~(right panel) for several models in Table~\ref{table_3}. The profiles are similar to those shown in Fig.~\ref{fig:rigid} of an isolated rotating white dwarf, but with a temporal scale of milliseconds instead of seconds. Actually, the final equilibrium value of $\Omega_c$ oscillates slightly around a stable value which, depending on the model, lays in the range $ 10^3$~s$^{-1} \le \Omega_c \le 10^4$~s$^{-1}$. This corresponds to periods of 0.6~ms~$\le P \le$ 6~ms, typical of millisecond pulsars. Figure~\ref{fig:NScolormap} shows several density colormaps along meridional slices of models B$_n$. The first column depicts the colormaps of pseudo-rigid rotators ($A=2$), being the case $m=1$ (model B$_1$) less elongated than $m=1/2$ (model B$_4$). According  to EM, this last case is very close to the critical value $R_p/R_{eq}=0.5988$, at which the gravitational and centrifugal forces become equal at the surface of the configuration when the rotation is $100\%$~rigid \citep{eri85}. 
 
 Models $B_3$ and $B_5$ ($A=0.2$) host a strong differential rotation and become highly deformed, with $\rho_{max}$ achieved far from the geometrical center of the configuration (see Fig.~\ref{fig:NScolormap}, right column). In particular, model B$_3$ is close to the limiting configuration at which our relaxation method gives satisfactory results when the trial initial configuration (prior relaxation) is a spherically symmetric LE model. This is because of the large contrast between the angular velocity at the center and at the surface equator, which in model B$_3$ is $\Omega_c/\Omega_{eq}\simeq 40$. Obtaining suitable models for much larger angular velocity ratios would require a more refined trial initial configuration, for instance those obtained with the SCF method, and increased resolution, which is left for a future work.
 
% FIGURE 5. EVOLUTION OF CENTRAL DENSITY AND OMEGA IN DIFFERENTIAL ROTATION. COLORMAP OF DENSITY WITH PROFILE OF OMEGA SUPERPOSED. 
\begin{figure*}
\centerline{\includegraphics[width=\textwidth]{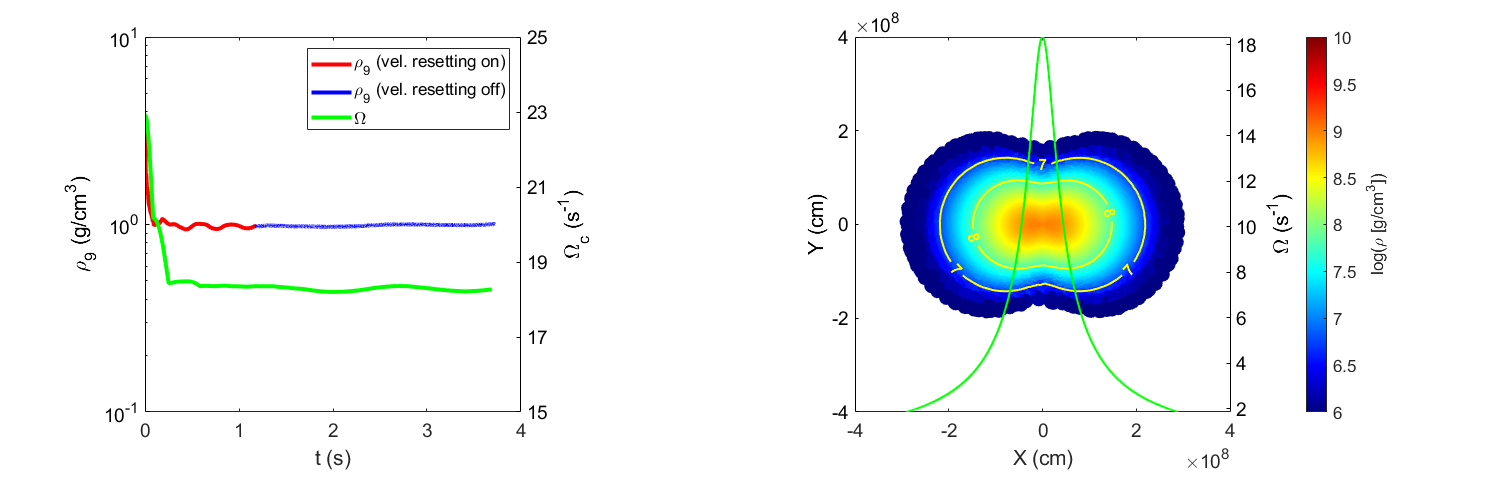}}
% where an .eps filename suffix will be assumed under latex,
% and a .pdf suffix will be assumed for pdflatex
\label{Differential}
\hfil
\caption{Time evolution of the maximum density and central angular velocity in model $A_7$, with differential rotation (left). Superposed to the density color map,  with thickness 4h, is the 1D profile of angular velocity (green) along a one-dimensional cut on the equator plane (right).}
\label{fig:differential}
\end{figure*}

 \subsection{Case 2: isolated WD with differential rotation}
 \label{subsec:differentialWD}
 We consider again a spinning zero-temperature white dwarf but, this time, with differential rotation. In particular, we have focused on initial models obeying  the rotation-law given by Eq.~(\ref{rotationlaw_1}) and $m=1/2$. This choice leads to rigid rotation at the center ($s\le R_c$) of the configuration, but becomes keplerian at distances $s >> R_c$. As previously stated, the case $m=1/2$ is usually referred as $v$-constant in the literature. As it can be seen in models A$_{6,7}$ in Table~\ref{table_1}, we find a good agreement among our models and those reported by Hachisu for the case $m=1/2$~(models H$_{6,7}$ in Table~\ref{table_1}). For these models $R_c = 0.1R_{eq}$, where the value of $R_{eq}$ was taken from the work by Hachisu (models 3 and 4 with $\rho_{max}=10^9$~g/cm$^3$ in their Table~4). Another example of differential rotation, with an exponent intermediate between $m=1/2$ and $m=1$, is discussed in the following section.

 We show a summary of our results for the two cases with $m=1/2$ in Table~\ref{table_1} and Fig.~\ref{fig:differential}. The two models differ from the SCF calculations in less than $5\%$, being stable enough for further hydrodynamic calculations. Figure~\ref{fig:differential} (left), shows the evolution of the maximum density and central value of the angular velocity for model $A_7$. At $t\simeq 1$~s both magnitudes, $\rho_{max}$ and $\Omega_c$, become stable. The profile of the angular velocity $\Omega$ has been superposed to the density color-map (right). It is clear that the maximum density is not located at the center of mass of the configuration, which is a typical signature of models with high angular momentum. The value of the angular velocity $\Omega$ is maximum at the center of the configuration and is very high, $\Omega (s=0)\simeq 18$~s$^{-1}$.
 
 After being relaxed, the rotating white dwarf is allowed to evolve freely during $\simeq 10$~complete orbits of the fastest particle. During that time the central density remains constant (left panel in Fig.~\ref{fig:differential}) while the central angular velocity $\Omega_c$~slightly oscillates around $18.4$~s$^{-1}$. As a matter of fact, what is shown in Fig.~\ref{fig:differential} at $t>1.2$~s is the averaged value of $\Omega_c$~for all particles with $s\ge 3\times 10^6$~cm. An estimate of $\Omega_c (s_a)$~is obtained from the tangential velocity of particle $a$,~$v_{t}(s_a,t)$,~through the Eq.~(\ref{rotationlaw_1}), 
 
 \begin{equation}
 \Omega_c (s_a,t) = \frac{v_{t}(s_a,t)}{s_a(t)}\times {\left(1 + \frac{s_a^2(t)}{R_c^2}\right)^m}
 \label{Omegac1}
 \end{equation}

%\subsection{Pseudo-Keplerian disks orbiting around a central mass-point}
\section{Pseudo-Keplerian disks orbiting a central mass-point}
\label{sec:disk}

Steady rotating disks are a universal phenomena in astrophysics \citep[e.g.][]{power} and often appear in merging stellar binary systems and during stellar and planetary formation \citep{armit2011}. Because of the thermal pressure contribution, these disks can not be described in a purely  Keplerian way. Pressure gradients effects are important and have to be taken into account to adequately model the structure of steady or pseudo-steady disks. Nevertheless, the simulation of pressure supported accretion disks with any hydrodynamic method has been proven difficult, especially if one wants to keep track the structure during many orbital periods. In the particular case of SPH codes, there is an extra difficulty in building a stable enough initial distribution of particles with pressure gradients in differential rotation \citep{owen2004}.

With the aim of checking the abilities of our relaxation scheme, we have implemented the generalized disk test problem discussed by \cite{ras16}. Such idealized disk was assumed to have cylindrical geometry and can be studied  in two-dimensions. As shown by these authors, the gravitational potential $\phi(r)$, pressure ($P(r)=K\rho^\gamma$, with $\gamma=3/2$), density, and tangential velocity ($v_\theta$) profiles follow precise analytic relationships which we reproduce here for completeness, 

\begin{equation}
\phi(r)=-\frac{GM}{(r^2+r_s^2)^\frac{1}{2}}\,
\label{potential}
\end{equation}

\begin{equation}
    \rho(r)=\left[\frac{GM(\gamma-1)}{K\gamma(r^2+r_s^2)^\frac{1}{2}}\right]^{\frac{1}{(\gamma-1}}
    \label{diskdens}
\end{equation}

The expressions above correspond to a pressure supported disk with zero angular momentum. To introduce  rotation, a reduction factor $f_p$ in the pressure is assumed so that the EOS is $P(r)=K\rho^\gamma f_p$. To keep the equilibrium, the loss in the pressure-gradient force is compensated by adding a centripetal force created by a tangential velocity field,  

\begin{equation}
v_\theta^2(r)=(1-f_p)\frac{GM r^2}{(r^2+r_s^2)^\frac{3}{2}}\,.
\label{veltan}
\end{equation}

For this test we set $GM=1, r_s=0.5$~and $f_p=0.5$. The value of $K$~in the EOS is obtained from Eq.~\ref{diskdens}, assuming ${\rho(r=0)=\rho_0=1}$, giving $K=2/3$. The resulting disk is pseudo-Keplerian with an important contribution of pressure. Finally, we want to study both, the resulting profiles of our relaxed initial models and the ability of SPHYNX to keep the disk in steady rotation during many orbits.

% FIGURE 6. PROFILES OF  DENSITY, TANGENTIAL VELOCITY AN %PRESSURE FOR THE DISK IN DIFFERENTIAL ROTATION. ALSO %SHOWN IS THE COLORMAP OF DENSITY IN THE CENTRAL REGION.   NOT YET DEFINITIVE (LACKS A ZOOM IN PRESSURE PROFILE AND IMPROVE THE CAOLORMAP) 

\begin{figure*}
\centerline{\includegraphics[width=\textwidth]{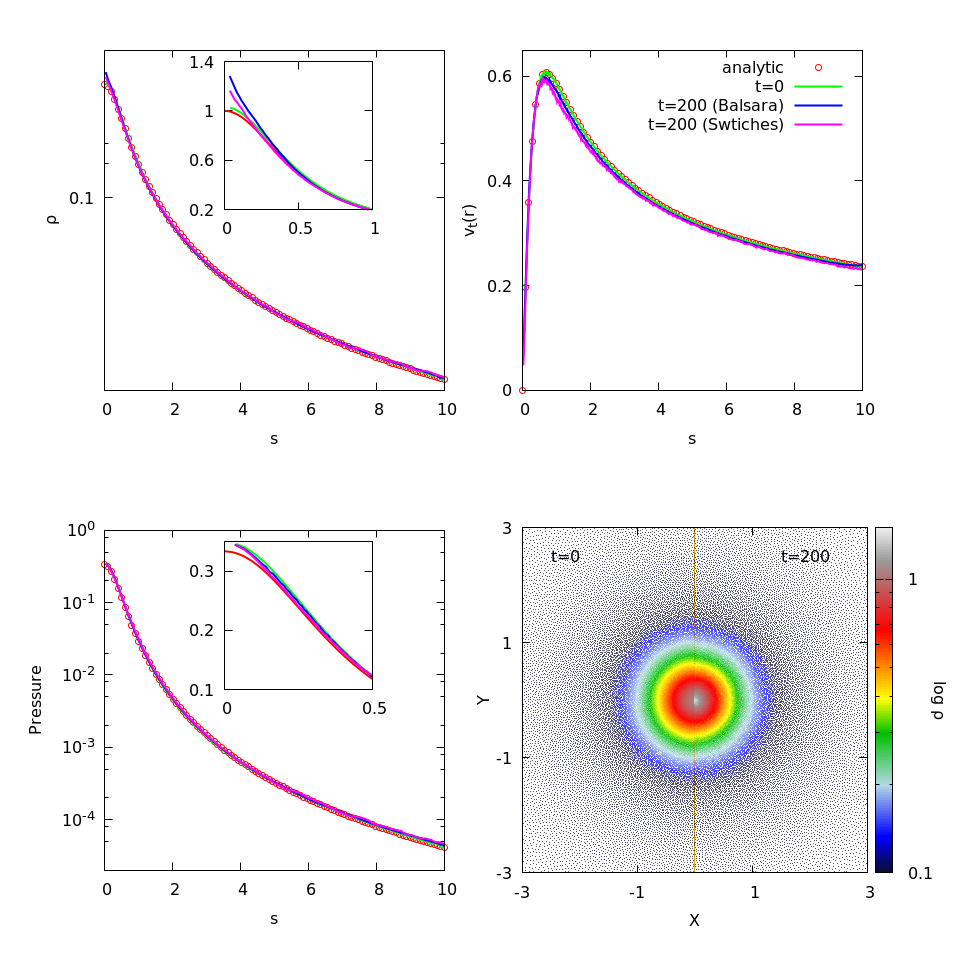}}
% where an .eps filename suffix will be assumed under latex,
% and a .pdf suffix will be assumed for pdflatex
\caption{Profiles of density, tangential velocity and pressure of the disk after the relaxation period (green lines) compared to the analytic values (in red) and after $t=200$~s calculated with the Balsara limiter (blue) or the AV switches by \cite{read12} (magenta). The bottom-right panel depicts the density color-map in the central region of the disk. The left semi-plane ($x < 0$) is for the relaxed model at $t=0$~s whereas the right semi-plane ($x\ge 0$)~shows the density color-map after $t= 200$~s, with the Balsara limiter. Note that,  unlike in \cite{ras16}, the SPH-particles are still settled in a glass-like configuration.}
\label{fig:disk}
\end{figure*}

Being a 2D calculation it is feasible to build an initial model by spreading the particles in an ordered array according to the density profile (Eq.~\ref{diskdens}). As shown by \cite{owen98} and \cite{ras16}, it is enough to distribute the particles in rings so that the radial separation between particles is adjusted in each annulus such that the angular separation between particles at a given radial coordinate is constant. In this work we propose a different procedure, which is capable to generate good initial models but with glass-like structure. 

Firstly, we fit the angular velocity $\Omega (r)= v_\theta (r)/r$ obtained from Eq.~(\ref{veltan}) with good accuracy, by the rotation law given by Eq.~(\ref{rotationlaw_1}) with $\Omega_c(t=0)=2$, $m=0.730$~and $r_c=0.478$. The total angular momentum of the disk, $J_D$, is estimated from the fitted $\Omega (r)$ and $\rho(r)$, with $0\le r\le R_D$, where $R_D=10$ is the adopted disk radius. The main features of our initial model are summarized in Table~\ref{table_4}. A sample of $N=9\times 10^4$ particles\footnote{Such particle count, spread in a disk radius $R_D=10$, is equivalent to the $N=7,800$ particles and $R_D=3$ adopted in \cite{ras16}, so that the comparison is meaningful.}  was radially distributed in a plane according to the density profile, while their angular position $0\le\theta \le2\pi$~is set at random. Afterwards, the particle sample is relaxed with the procedure described in Section~\ref{sec:relax}.

\begin{table}
%%% increase table row spacing, adjust to taste
\renewcommand{\arraystretch}{1.3}
\caption{Main features of the disk model after being relaxed: mass, angular momentum, rotation constants: $\Omega_c [s^{-1}]$, $R_c$~and $m$~in Equation \ref{rotationlaw_1}, and radius.}
\label{table_4}
\begin{center}
%%% Some packages, such as MDW tools, offer better commands for making tables
%%% than the plain LaTeX2e tabular which is used here.
\begin{tabular}{|c|c|c|c|c|c|c|c|c|}
\hline
%\bf{Model} & \multicolumn{3}{|c|}{\bf{Table Column Head}}\\
\bf{}
%\cline{2-4}
  {M$_{D}$}&J$_D$& $\Omega_c$ & R$_c$ & $m$&R$_D$  \\
\hline
\hline
4.7076 & 5.0298 & 2.004358 &0.478  &0.73&10.057 \\
\hline
\end{tabular}
\end{center}
\end{table}

Figure~\ref{fig:disk} shows the density, pressure and tangential velocity profiles of the particles once the relaxation has ended ($t=0$ cases). A density color-map showing the central region of the disk with $r\le 3$~is depicted in the left semi-plane of the bottom-right panel in the same figure. In spite of the disordered, glass-like pattern, the distribution of the SPH-particles in the disk is matching the analytic profiles very well. But this is not the end of the history because, after the relaxation, such pseudo-Keplerian  disk  should retain its structure  during as many orbits as possible. It was shown by \cite{ras16} that many current SPH schemes fail to preserve the disk features after some dozens of periods of the particle with maximum velocity. Satisfactory results were obtained with the conservative reproducing kernel (CRKSPH) method by \cite{front18} in combination with the AV switches by \cite{cullen10}.   

The stability of the disk obtained with our relaxation scheme was further checked with the SPHYNX code. We want to know if the integral approach to the gradients in combination with either the Balsara limiter \citep{balsara95} or the AV switches \citep{cullen10,read12}, is capable to keep the disk in steady state during many orbital periods. For this calculation the EOS was changed to $P=(\gamma-1)\rho u$~($\gamma=5/3$) where $u$~is the specific internal energy of the gas. The evolution of the internal energy was obtained by evolving the corresponding energy equation. 

We have tracked the evolution until $t=200$~s, so that the results can be compared to those obtained by \cite{ras16} with the CRKSPH method. At $t=200$~s the particle of highest velocity has completed $\simeq 40$ orbits around the center of the disk. As shown in Fig.~\ref{fig:disk}, the deviation of $\rho(s), v_t (s)$, and $P(s)$ from either the analytic or the $t=0$~s profiles is very small. Particularly good is the fit of the tangential velocity, which is still matching the initial profile after $\simeq 40$~orbits. Nevertheless, a closer look into the density profile in the central region indicates that its value is slowly growing with time. We agree with  \cite{ras16}, that such small growth of the central density is an artifact of the particular implementation of the artificial viscosity. It is slightly less pronounced when the switches to the AV are in command of the dissipation. It is worth noting, however, that the pressure profile remains very close to the analytic value.

The $x\ge 0$~semi-plane in the bottom-right panel in Fig.~\ref{fig:disk} depicts the density colormap at $t=200$~s in the central part of the disk. As we can see, the differences with the colormap at $t=0$~s ($x < 0$~semi-plane) are small, being only relevant at the very center of the configuration. Additionally, the model is still retaining a good glass-like granulation, with no indications of pairing-instability after $\simeq 40$~revolutions. 

\section{Conclusions}
\label{sec:conclusions}

A common problem in simulating the evolution of compact objects with SPH is that there is not a general procedure to obtain stable initial models when these objects are spinning fast. In this paper, we propose and test an easy and versatile relaxation scheme to build stable rotating models of self-gravitating bodies whose EOS is of barotropic type. The hypothesis of barotropic EOS allows to handle many interesting objects such as isothermal white dwarfs, neutron stars, and polytropic structures. As detailed in Sec.~\ref{sec:relax}, our method relies in the exceptional angular momentum conservation properties of the SPH technique.

We apply our method to get stable rotating configurations of zero-temperature WDs with different masses and total angular momentum. We were able to build stable models of rotating white dwarfs and polytropes spanning a wide mass-range, $0.67M_{\odot}\le M_{WD}\le 2 M_{\odot}$, with both, rigid and  differential rotation. The main magnitudes: central density, total kinetic, internal and gravitational energies, equatorial and polar radius, match the semi-analytical results by \cite{hach86} and \cite{eri85}, in general within a few percents. Given the current uncertainties in the particular rotation-law followed by these compact objects, that precision is enough to explore many issues concerning to their evolution, either for explosion or collapse scenarios. Additionally, we show that our method is able to produce stable configurations when it is applied to a pair of white dwarfs orbiting in a compact binary system. This last scenario is of capital importance to understand the double-degenerate route to Type Ia supernova explosions. 

Finally, we have applied the method to produce a steady accretion disk with cylindrical symmetry. Even tough the assumption of cylindrical geometry is not realistic, such configuration has the advantage that it has an analytical solution to compare with, while still retaining many of the features of real disks. The ensuing relaxed disk is characterized by a glass-like distribution of particles in differential rotation, partially supported by pressure effects and by rotation. As shown in Sec.~\ref{sec:disk}, such pseudo-keplerian disk is in a good steady state, being stable during at least several dozens of orbits of the particle with maximum velocity.           

Presently, two cautionary remarks have to be taken into account before applying our procedure to astrophysical calculations. First, the centrifugal force has to remain smaller than the gravitational pulling at the equator when approaching the equilibrium configuration. Otherwise a harmful trade-off between gravity ($\propto 1/s^2$)~and the centrifugal force ($\propto s)$~ may appear which finally leads to the ejection of particles located in the outer shells along the equatorial plane. Second, in differential rotation the ratio between the angular velocity at the center and at the surface, prior relaxation, should be not too extreme. In particular, the combination of a shellular-like  ($m\simeq 1$) rotation with a high total angular momentum  and low values of $R_c$~in Eqs. (\ref{rotationlaw_1}) and (\ref{momentumcons_2}) lead to toroidal structures where $r_p/r_{eq}\rightarrow 0$~\citep{eri85}.  Troubles may also appear when trying to relax bodies with extremely large density contrasts between the core and the surface layers, owing to their very different characteristic time-scales. In these cases, the  choice of spherically symmetric Lane-Endem models as initial trial configurations is not adequate because they are too far from the final equilibrium structure and  the relaxation to a steady state could be hard or even impossible.  Such difficulty can probably be overcomed using better initial trial models, as for example those obtained with the SCF method or even envisaging the final toroidal structure as the result of the coalescence of two  spherically symmetric objects (as it was done in Section \ref{sec:DD}). 

The practical cases studied in this work cover only a small subset of the potential applications. An immediate additional application of our method may consist in building 3D stable post-Newtonian models of fast-rotating neutron stars, either isolated or in binary systems. That task should not be difficult because the equation of state of cold neutron stars can also be described assuming zero temperature. 

Prospects for future extensions of this work may include the impact of finite temperature gradients in the relaxed rotating structures. Actually, the steady spinning polytropes obtained in Section \ref{subsec:differentialNS} with a barotropic EOS, $P=K\rho^\gamma$, can easily be mapped into an axis-symmetric distribution of temperature if we assume a realistic EOS, $P(\rho, T)$ (f.e. that of a non-degenerate ideal gas). This suggests that relaxing a trial initial model with a realistic EOS besides a superimposed axis-symmetric temperature profile, $T(s)$, from the onset is feasible.

\section*{Acknowledgments}
This work has been supported by the MINECO Spanish project AYA2017-86274-P and the Generalitat of Catalonia SGR-661/2017 (DG), and by the Swiss Platform for Advanced Scientific Computing (PASC) project SPH-EXA: Optimizing Smooth Particle Hydrodynamics for Exascale Computing (RC and DG). JMBI acknowledges the support by the FPU fellowship and wants to thank the Ministerio de Educaci\'on, Cultura y Deporte from Spain. The authors acknowledge the support of sciCORE (http://scicore.unibas.ch/) scientific computing core facility at University of Basel, where part of these calculations were performed.

%\bibliographystyle{aa}
%\bibliography{bibliography_r} 

\end{document}